\documentclass[pra,aps,preprint,showpacs]{revtex4}
\usepackage{txfonts}
\usepackage{amssymb}
\newcommand{\ket}[1]{|#1\rangle}
\newcommand{\bra}[1]{\langle #1|}

\begin{document}
\title{Separable states and the geometric phases of an interacting two-spin system}
\author{C. W. Niu,  G. F. Xu, Longjiang Liu, L. Kang and D. M.
Tong\footnote{tdm@sdu.edu.cn}} \affiliation{Department of Physics,
Shandong University, Jinan 250100, China \\}
\author{L. C. Kwek} \affiliation{Center
for Quantum Technologies, National University of Singapore, Science Drive 2
Singapore 117543}
\affiliation{Institute of Advanced Studies, Nanyang Technological University, 60
Nanyang View Singapore 639673}

\date{\today}
\begin{abstract}
It is known that an interacting bipartite system evolves as an
entangled state in general, even if it is initially in a separable
state. Due to the entanglement of the state, the geometric phase of
the system is not equal to the sum of the geometric phases of its
two subsystems. However, there may exist a set of states in which
the nonlocal interaction does not affect the separability of the
states, and the geometric phase of the bipartite system is then
always equal to the sum of the geometric phases of its subsystems.
In this paper, we illustrate this point by investigating a well
known physical model. We give a necessary and sufficient condition
in which a separable state remains separable so that the geometric
phase of the system is always equal to the sum of the geometric
phases of its subsystems. 
\end{abstract}
\maketitle
\date{\today}

\section{Introduction}

The notion of geometric phase was first addressed by Pancharatnam
for the comparison of the phases of two beams of polarized light in 1956
\cite{pancharatnam}. It was later shown to have important
consequences for quantum systems. In 1984, Berry demonstrated that
quantum system undergoing a cyclic adiabatic evolution acquires a
phase with geometric nature \cite{Berry}. Since then, geometric
phase has attracted great interest. The original notion of Berry
phase has been extended to nonadiabatic cyclic evolution by Aharonov
and Anandan in 1987 \cite{Aharonov}, and to nonadiabatic and
noncyclic evolution by Samuel and Bhandari in 1988 \cite{Samuel}.

While all these extensions of quantum systems are in pure states,
another line of development has been towards extending the geometric
phase to mixed states. The early extension to mixed states was given
by Uhlmann within the mathematical context of purification
\cite{Uhlmann}. In 2000, Sj$\ddot{o}$qvist et al. introduced an
alternative definition of geometric phases for mixed states under
unitary evolution based on quantum interferometry \cite{Sj}, and
subsequently Singh et al. gave a kinematic description of the mixed
state geometric phase and extended it to degenerate density operator
\cite{Singh}. The generalization of mixed geometric phases to
quantum systems in nonunitary evolution was given by Tong et al. in
2004 \cite{Tong}. Other discussions or experimental demonstrations
of geometric phases for mixed states may be found in papers
\cite{Carollo,Ericsson,Marzlin,Kamleitner,Whitney,yi,Lombardo,Bassi,Sarandy,Rezakhani,Goto,Mller,Buric,Du,
Ericsson2,yin}.

Another interesting issue of geometric phase is the relation of
the bipartite or multipartite system with its subsystems.
Sj$\ddot{o}$qvist calculated  the geometric phase of a pair of
entangled spin half particles precessing in a time-independent
uniform magnetic field \cite{Sj1}, and the relative phase for
polarization-entangled two-photon systems was considered by Hessmo
et al \cite{Hessmo}. Tong et al. calculated the geometric phase of
a bipartite entangled spin-half system in a rotating magnetic field
\cite{Tong2} and investigated entangled bipartite systems with local
unitary evolutions \cite{Tong1}.  The effect of entanglement on the
mutual geometric phase was recently studied by Williamson et al
\cite{Williamson}.  Other discussions on geometric phases of
composite systems and its applications may be found in Refs
\cite{yi1,Li,yi3,Ge,Xing,wangxb}.

All the previous discussions concerning the relation of the
geometric phase of the composite system with its subsystems were of
the systems under local unitary evolutions, $U(t)=U_a(t)\otimes
U_b(t)$. It was shown that the geometric phase of the composite
system, $\gamma_{ab}$, does not equal the sum of the geometric
phases of its subsystems, $\gamma_a$ and $\gamma_b$, in general
\cite{Tong1,Williamson}. The expression
$\gamma_{ab}=\gamma_a+\gamma_b$ is valid only if the initial state
is a separable one. This is because that the entanglement of the
state leads to an indecomposable geometric phase of the composite
system. Since the interaction between two subsystems can lead to an
entanglement of the subsystems, it is usually deemed that the
geometric phase of the composite system in nonlocal unitary
evolution does not equal the sum of the geometric phases of its
subsystems in general, even if the initial state of the system is
separable. In the present paper, we investigate a well known
physical model, two interacting spin-half particles in a rotating
magnetic field.  We aim to show that there may exist a set of states
in which the nonlocal interaction does not affect the separability
of the states, and therefore the geometric phase of the bipartite
system is always equal to the sum of the geometric phases of its
subsystems. A necessary and sufficient condition for the set of
separable states is given.

\section{The interacting two-spin half model}
Consider the system of two interacting spin-half particles in a
rotating magnetic field, the Hamiltonian of which is described as
\begin{eqnarray}
\hat{H}(t)=\hat{H}_a(t)\otimes
I+I\otimes\hat{H}_b(t)+\hat{H}_{ab}(t),
\end{eqnarray}
where $\hat{H_\mu}(t)=\vec{B}(t)\cdot\vec{\sigma}_\mu$ $~(
\mu=a,b)$, $ \hat{H}_{ab}(t)=J\vec{\sigma}_a\cdot\vec{\sigma}_b$.
Here, $\vec{B}(t)=B(\sin\theta\cos\omega t, \sin\theta\sin\omega t,
\cos\theta)$ is the rotating magnetic field. $\vec{\sigma}_a$ and
$\vec{\sigma}_b$ are the Pauli operators of spins $a$ and $b$,
respectively. $J$ denotes the interaction strength between $a$ and
$b$, and  $J>0$ describes antiferromagnetic coupling and $J<0$
describes ferromagnetic coupling.

The state of the system, $\ket{\psi(t)}$ , satisfies the
Schr$\ddot{o}$dinger equation,
\begin{eqnarray}
i\frac{d}{dt}\ket{\psi(t)}=\hat{H}(t)\ket{\psi(t)}, \label{a1}
\end{eqnarray}
with initial state being $\ket{\psi(0)}$.  $\ket{\psi(t)}$ may be
expressed as
\begin{eqnarray}
\ket{\psi(t)}=f_1(t)\ket{00}+f_2(t)\ket{01}+f_3(t)\ket{10}+f_4(t)\ket{11},
\label{a2}
\end{eqnarray}
where $\ket{ij}$ $ (i,j=0,1)$ are the abbreviations of
$\ket{i}\otimes\ket{j}$ with $\ket{0}=\bordermatrix{&
\cr&1\cr&0\cr}$ and $\ket{1}=\bordermatrix{& \cr&0\cr&1\cr}$, and
$f_k(t)~~(k=1,2,3,4)$ are functions of $t$ to be determined,
satisfying $\sum_{k=1}^{4}|f_{k}(t)|^{2}=1$. Substituting
Eq.(\ref{a2}) into Eq. (\ref{a1}), we have
\begin{eqnarray}
i\frac{d}{dt}\bordermatrix{&\cr &f_1(t) \cr &f_2(t) \cr &f_3(t)\cr
&f_4(t)\cr}=\bordermatrix{&\cr &J+2B\cos\theta &B\sin\theta
e^{-i\omega t} &B\sin\theta e^{-i\omega t} &0 \cr &B\sin\theta
e^{i\omega t} &-J &2J &B\sin\theta e^{-i\omega t} \cr &B\sin\theta
e^{i\omega t} &2J &-J &B\sin\theta e^{-i\omega t}\cr &0&B\sin\theta
e^{i\omega t}&B\sin\theta e^{i\omega t}&J-2B\cos\theta\cr}
\bordermatrix{&\cr &f_1(t)\cr &f_2(t)\cr &f_3(t)\cr &f_4(t)\cr},
\end{eqnarray}
that is,
\begin{equation}
\left\{
\begin{array}{rl}
i \dot{f}_1&=(J+2B\cos\theta) f_1+(B\sin\theta)e^{-i\omega t}f_2+(B\sin\theta)e^{-i\omega t}f_3,\\
i \dot{f}_2&=(B\sin\theta)e^{i\omega t}f_1-Jf_2+2Jf_3+(B\sin\theta)e^{-i\omega t}f_4,\\
i \dot{f}_3&=(B\sin\theta)e^{i\omega t}f_1+2Jf_2-Jf_3+(B\sin\theta)e^{-i\omega t}f_4,\\
i \dot{f}_4&=(B\sin\theta)e^{i\omega t}f_2+(B\sin\theta)e^{i\omega
t}f_3+(J-2B\cos\theta)f_4.
\end{array}
\right.\label{eqs}
\end{equation}
To resolve the above differential equations,  we further let
$f_1(t)=\bar{f}_1(t)e^{-i\omega
t},~~f_2(t)=\bar{f}_2(t),~~f_3(t)=\bar{f}_3(t),~~f_4(t)=\bar{f}_4(t)e^{i\omega
t}$. Then,  Eq.(\ref{eqs}) becomes
\begin{equation}
\left\{
\begin{array}{rl}
i\dot{\bar{f}}_1&=(J+2B\cos\theta-\omega)\bar{f}_1+B\sin\theta \bar{f}_2+B\sin\theta \bar{f}_3,\\
i\dot{\bar{f}}_2&=B\sin\theta \bar{f}_1-J\bar{f}_2+2J\bar{f}_3+B\sin\theta \bar{f}_4,\\
i\dot{\bar{f}}_3&=B\sin\theta \bar{f}_1+2J\bar{f}_2-J\bar{f}_3+B\sin\theta \bar{f}_4,\\
i\dot{\bar{f}}_4&=B\sin\theta \bar{f}_2+B\sin\theta
\bar{f}_3+(J-2B\cos\theta+\omega)\bar{f}_4.
\end{array}
\right.\label{equation1}
\end{equation}
Eq. (\ref{equation1}) is a set of first-order linear ordinary
differential equations. Its solution can be obtained by solving the
characteristic equation. The four characteristic roots are
\begin{eqnarray}
\lambda_1 &=& 3J,\nonumber\\
\lambda_2 &=&-J,\nonumber\\
\lambda_3 &=&-J+\sqrt{4B^2\sin^{2}\theta+(2B\cos\theta-\omega)^{2}},\nonumber\\
\lambda_4 &=&-J-\sqrt{4B^2\sin^{2}\theta+(2B\cos\theta-\omega)^{2}},
\end{eqnarray}
 each of which corresponding to a characteristic
solution with respect to $\bar{f}_{k}(t)$. With the help of the
solutions of $\bar{f}_k(t)$, which directly give the solutions of
$f_k(t)$, the general solution of Eq.({\ref{a1}}) can be expressed
as
\begin{eqnarray}
\ket{\psi(t)}=c_1\ket{\psi_{1}(t)}+c_2\ket{\psi_{2}(t)}+c_3\ket{\psi_{3}(t)}+c_4\ket{\psi_{4}(t)},\label{gn}
\end{eqnarray}
where the time-independent coefficients $c_k$ $(k=1,2,3,4)$,
$\sum_{k=1}^4|c_k|^2=1$, are to be determined by the initial
condition, and the four particular solutions read
\begin{eqnarray}
\ket{\psi_1(t)}&=&e^{i\lambda_1t}\frac{1}{\sqrt{2}}\bordermatrix{&\cr
&0 \cr &1  \cr &-1 \cr &0 \cr},\nonumber\\
\ket{\psi_2(t)}&=&e^{i\lambda_2t}\frac{1}{\sqrt{2}}\bordermatrix{&\cr
&-\frac{2B\sin\theta}{\sqrt{4B^2\sin^2\theta+(2B\cos\theta-\omega)^2}}e^{-i\omega
t} \cr
&\frac{2B\cos\theta-\omega}{\sqrt{4B^2\sin^2\theta+(2B\cos\theta-\omega)^2}}
\cr
&\frac{2B\cos\theta-\omega}{\sqrt{4B^2\sin^2\theta+(2B\cos\theta-\omega)^2}}
\cr
&\frac{2B\sin\theta}{\sqrt{4B^2\sin^2\theta+(2B\cos\theta-\omega)^2}}e^{i\omega
t}
\cr},\nonumber\\
\ket{\psi_3(t)}&=&e^{i\lambda_3t}\bordermatrix{&\cr
&-\frac{\sqrt{4B^2\sin^{2}\theta+(2B\cos\theta-\omega)^2}-(2B\cos\theta-\omega)}{2\sqrt{4B^2\sin^2\theta+(2B\cos\theta-\omega)^2}}e^{-i\omega
t}\cr
&\frac{B\sin\theta}{\sqrt{4B^2\sin^2\theta+(2B\cos\theta-\omega)^2}}\cr
&\frac{B\sin\theta}{\sqrt{4B^2\sin^2\theta+(2B\cos\theta-\omega)^2}}\cr
&-\frac{\sqrt{4B^2\sin^{2}\theta+(2B\cos\theta-\omega)^2}+(2B\cos\theta-\omega)}
{2\sqrt{4B^2\sin^2\theta+(2B\cos\theta-\omega)^2}}e^{i\omega t}\cr},\nonumber\\
\ket{\psi_4(t)}&=&e^{i\lambda_4t}\bordermatrix{&\cr
&\frac{\sqrt{4B^2\sin^{2}\theta+(2B\cos\theta-\omega)^2}+(2B\cos\theta-\omega)}{2\sqrt{4B^2\sin^2\theta+(2B\cos\theta-\omega)^2}}e^{-i\omega
t}\cr
&\frac{B\sin\theta}{\sqrt{4B^2\sin^2\theta+(2B\cos\theta-\omega)^2}}\cr
&\frac{B\sin\theta}{\sqrt{4B^2\sin^2\theta+(2B\cos\theta-\omega)^2}}\cr
&\frac{\sqrt{4B^2\sin^{2}\theta+(2B\cos\theta-\omega)^2}-(2B\cos\theta-\omega)}{2\sqrt{4B^2\sin^2\theta+(2B\cos\theta-\omega)^2}}e^{i\omega
t}\cr}.\label{a3}
\end{eqnarray}

\section{The geometric phases of the two-spin half system}
If the two-spin half system is initially in state $\ket{\psi(0)}$,
the geometric phase obtained by the quantum system during the time
$t\in[0,\tau]$ can be calculated by using the formula
\cite{Bhandari,Mukunda},
\begin{eqnarray}
\gamma_{ab}(\tau)=\arg\langle{\psi(0)}\ket{\psi(\tau)}+i\int_{0}^{\tau}\langle{\psi(t)}\ket{\dot{\psi}(t)}dt.\label{pure}
\end{eqnarray}
However, both the subsystems $a$ and $b$ are generally in mixed
states due to the nonlocal interaction, even if the initial state
$\ket{\psi(0)}$ is separable. The mixed states of the subsystems can be
expressed as density operators,
\begin{eqnarray}
\rho_a(t)=tr_b\ket{\psi(t)}\bra{\psi(t)},~~\rho_b(t)=tr_a\ket{\psi(t)}\bra{\psi(t)}.
\label{rhoab}
\end{eqnarray}
The geometric phases of the mixed states in nonunitary evolutions
are calculated by using the formula \cite{Tong}
\begin{eqnarray}
\gamma_\mu(\tau)=\arg\left(\sum_{m=1}^{2}\sqrt{\omega^\mu_{m}(0)\omega^\mu_{m}(\tau)}\langle{\phi^\mu_{m}(0)}\ket{\phi^\mu_{m}(\tau)}
e^{-\int_{0}^{\tau}\langle{\phi^\mu_{m}(t)}\ket{\dot{\phi^\mu}_{m}(t)}dt}\right),
\label{mixed}
\end{eqnarray}
where $\omega^\mu_{m}(t)$ and $\ket{\phi^\mu_{m}(t)}$ are the
eigenvalues and eigenstates of the density operators $\rho_\mu(t)$ (
$\mu=a,b$), respectively.

By substituting Eqs. (\ref{gn}) and (\ref{a3}) into Eqs.
(\ref{pure}) and (\ref{rhoab}), and further using Eq. (\ref{mixed}),
one can calculate the geometric phase of the two-spin half system
and the geometric phases of its two subsystems. It is easy to show
that $\gamma_{ab}$ is not equal to the sum of $\gamma_a$ and
$\gamma_b$ in general, even if the initial state $\ket{\psi(0)}$ is
a separable one.

To illustrate this point, we take $\ket{\psi(0)}=\ket{01}$ as an
example. In this case, the state of the system at time $t$ reads
\begin{eqnarray}
\ket{\psi(t)}=\frac{1}{\sqrt{2}}\ket{\psi_{1}(t)}+\frac{1}{\sqrt{2}}\cos\eta
\ket{\psi_{2}(t)}+\frac{1}{2}\sin\eta
\ket{\psi_{3}(t)}+\frac{1}{2}\sin\eta \ket{\psi_{4}(t)},
\end{eqnarray}
and the geometric phase obtained by the system during the time
$t\in[0,\tau]$ is
\begin{eqnarray}
\gamma_{ab}=\arctan\frac{\sin4J\tau}{\cos^{2}\eta+\sin^{2}\eta\cos\alpha\tau+\cos4J\tau}-2J\tau,
\end{eqnarray}
where
\begin{eqnarray}
\alpha=\sqrt{4B^2\sin^{2}\theta+(2B\cos\theta-\omega)^{2}},\label{alpha}
\end{eqnarray}
and
\begin{eqnarray}
\tan\eta=\frac{2B\sin\theta}{2B\cos\theta-\omega}.\label{eta}
\end{eqnarray}
The reduced density operators of the subsystems $a$ and $b$ are
\begin{eqnarray}
\rho_\mu=\bordermatrix{&\cr &\rho^\mu_{11}&\rho^\mu_{12}\cr
&\rho^\mu_{21} &\rho^\mu_{22}},~~\mu=a,b,
\end{eqnarray}
where
\begin{eqnarray}
\rho^a_{11}&=&1-\rho^a_{22}=\frac{1}{2}\left(1+(\cos^2\eta+\sin^2\eta\cos\alpha
t)\cos4Jt\right),\nonumber\\
\rho^a_{12}&=&{\rho^{a}_{21}}^*=\frac{1}{2}\left(\sin\eta\cos\eta(1-\cos\alpha
t)+i\sin\eta\sin\alpha t\right)e^{-i\omega t} \cos4Jt;\\
\rho^b_{11}&=&1-\rho^b_{22}=\frac{1}{2}\left(1-(\cos^2\eta+\sin^2\eta\cos\alpha
t)\cos4Jt\right),\nonumber\\
\rho^b_{12}&=&{\rho^{b}_{21}}^*=-\frac{1}{2}\left(\sin\eta\cos\eta(1-\cos\alpha
t)+i\sin\eta\sin\alpha t\right)e^{-i\omega t} \cos4Jt.
\end{eqnarray}
The geometric phases obtained by the subsystems during the time
$t\in[0,\tau]$ are respectively
\begin{eqnarray}
\gamma_a(\tau)=\arctan\frac{-\cos\eta\sqrt{1-\cos\alpha\tau}}{\sqrt{1+\cos\alpha\tau}}
+\frac{\omega\sin^{2}\eta}{2\alpha}\sin\alpha\tau+\frac{1}{2}\alpha\tau\cos\eta
-\frac{1}{2}\omega\tau\sin^{2}\eta,\label{g01}
\end{eqnarray}
and
\begin{eqnarray}
\gamma_b(\tau)=\arctan(\cos\eta\tan\frac{\alpha\tau}{2})-\frac{\omega\sin^{2}\eta}{2\alpha}\sin\alpha\tau-\frac{1}{2}\alpha\tau\cos\eta+\frac{1}{2}\omega\tau\sin^{2}\eta.\label{g01b}
\end{eqnarray}
Clearly, the geometric phase of the large system is not equal to the
sum of the geometric phases of the two subsystems, $\gamma_{ab}\neq
\gamma_{a}+ \gamma_{b}$, even if the initial state $\ket{\psi(0)}$
is a separable one.

\section{Condition for geometric phase of the system being equal to the sum of those of its subsystems}
Geometric phase is useful in quantum calculation, but a real quantum
system may comprise two or more subsystems with interactions between
them. In this case when interactions appear, the geometric phase of
the composite system is not equal to the sum of the geometric phases
of its subsystems. The relations among the geometric phases of the
large system and the subsystems are complicated, and therefore they
are not easy to be synchronously controlled. It is interesting to
find a condition in which the geometric phase of the composite
system  equals the sum of the geometric phases of its
subsystems\cite{tong3}. The formulae (\ref{pure}) and (\ref{mixed})
show that the value of geometric phase of a quantum system not only
depends on the initial state $\ket{\psi(0)}$ and the final state
$\ket{\psi(\tau)}$ but also depends on all the instantaneous states
$\ket{\psi(t)}~(t\in[0,\tau])$. It is determined completely by the
path traced by the states. If we require that the geometric phase of
the composite system is equal to the sum of the geometric phases of
its subsystems for all time, the sufficient condition is that
$\ket{\psi(t)}$ remains separable at all time, i.e.
\begin{eqnarray}
\ket{\psi(t)}=\ket{\phi_a(t)}\otimes \ket{\phi_b(t)}.\label{psiab}
\end{eqnarray}
One may demonstrate this point by substituting expression
(\ref{psiab}) into geometric phase formulae. Indeed, if there is
$\ket{\psi(t)}=\ket{\phi_a(t)}\otimes \ket{\phi_b(t)}$ for $t \in
[0,\tau]$, one then has
\begin{eqnarray}
\arg\langle{\psi(0)}\ket{\psi(\tau)}&=&\arg\langle{\phi_a(0)}\ket{\phi_a(\tau)}\langle{\phi_b(0)}\ket{\phi_b(\tau)}\nonumber\\
&=&\arg\langle{\phi_a(0)}\ket{\phi_a(\tau)}+\arg\langle{\phi_b(0)}\ket{\phi_b(\tau)}~~(\text{mod}
2\pi),
\end{eqnarray}
and
\begin{eqnarray}
i\int_{0}^{\tau}\langle{\psi(t)}\ket{\dot{\psi}(t)}dt=i\int_{0}^{\tau}\langle{\phi_a(t)}\ket{\dot{\phi}_a(t)}dt+i\int_{0}^{\tau}\langle{\phi_b(t)}\ket{\dot{\phi}_b(t)}dt,
\end{eqnarray}
where the normalized relations
$\langle\phi_\mu(t)\ket{\phi_\mu(t)}=1~ (\mu=a,b)$ are used.
Substituting them into Eq. (\ref{pure}), one further has
\begin{eqnarray}
\gamma_{ab}(\tau)= \gamma_{a}(\tau)+\gamma_{b}(\tau),
\end{eqnarray}
where
\begin{eqnarray}
\gamma_{\mu}(\tau)=\arg\langle{\phi_\mu(0)}\ket{\phi_\mu(\tau)}+i\int_{0}^{\tau}\langle{\phi_\mu(t)}\ket{\dot{\phi}_\mu(t)}dt,~~\mu=a,b,
\end{eqnarray}
are $2\pi$-modular geometric phases of the subsystems.

With the above knowledge that the geometric phase of the system is
equal to the sum of those of its subsystems if the time dependent
state is always separable, we may now calculate the condition for
the two interacting spin-half particles. To this end, we rewrite the
general solution expressed  by Eq. (\ref{gn}), with the bases
$\{\ket{00},\ket{01},\ket{10},\ket{11}\}$,  as
\begin{eqnarray}
\ket{\psi(t)} =&&\left(-\frac{1}{\sqrt{2}}c_{2}\sin\eta
e^{i\lambda_{2}t}e^{-i\omega
t}-c_{3}\sin^{2}\frac{\eta}{2}e^{i\lambda_{3}t}e^{-i\omega
t}+c_{4}\cos^{2}\frac{\eta}{2}e^{i\lambda_{4}t}e^{-i\omega
t}\right)\ket{00}\nonumber\\
&&+\left(\frac{1}{\sqrt{2}}c_{1}e^{i\lambda_{1}t}+\frac{1}{\sqrt{2}}c_{2}\cos\eta
e^{i\lambda_{2}t}+\frac{1}{2}c_{3}\sin\eta
e^{i\lambda_{3}t}+\frac{1}{2}c_{4}\sin\eta
e^{i\lambda_{4}t}\right)\ket{01}\nonumber\\
&&+\left(-\frac{1}{\sqrt{2}}c_{1}e^{i\lambda_{1}t}+\frac{1}{\sqrt{2}}c_{2}\cos\eta
e^{i\lambda_{2}t}+\frac{1}{2}c_{3}\sin\eta
e^{i\lambda_{3}t}+\frac{1}{2}c_{4}\sin\eta
e^{i\lambda_{4}t}\right)\ket{10}\nonumber\\
&&+\left(\frac{1}{\sqrt{2}}c_{2}\sin\eta e^{i\lambda_{2}t}e^{i\omega
t}-c_{3}\cos^{2}\frac{\eta}{2}e^{i\lambda_{3}t}e^{i\omega
t}+c_{4}\sin^{2}\frac{\eta}{2}e^{i\lambda_{4}t}e^{i\omega
t}\right)\ket{11}.
\end{eqnarray}

Noting that the concurrence of a quantum state provides a criterion
for distinguishing between separable states and entangled states
\cite{Wootters,Rungta}, we may obtain the necessary and sufficient
condition for the separable states by calculating the concurrence of
the above state. The concurrence of the state reads
\begin{eqnarray}
C(t)&=&\sqrt{2\left[1-tr(tr_b\ket{\psi(t)}\bra{\psi(t)})^2\right]}\nonumber\\
&=&\left|c_{2}^{2}+2c_{3}c_{4}-c_{1}^{2}e^{i8Jt}\right|.
\end{eqnarray}
The above equation shows that the concurrence is, if $J\neq 0$,
dependent on the time $t$. The separability of an initial state does
not guarantee separability of the state at time $t$. If the system
is initially in a separable state, satisfying $
c_{2}^{2}+2c_{3}c_{4}-c_{1}^{2}=0$ with $c_1\neq 0$, it will evolve
to an entangled state at the late time and then go back to a
separable state at each time $t=n\pi\left/4J\right.,~n=1,2,\cdots$.
If we require that the geometric phase of the composite system is
always equal to the sum of the geometric phases of its subsystems
for all time, the sufficient condition is that $\ket{\psi(t)}$ is
separable for all time. This requirement is fulfilled if and only if
the concurrence $C(t)$ is zero for all time, i.e.
\begin{eqnarray}
c_{2}^{2}+2c_{3}c_{4}-c_{1}^{2}e^{i8Jt}=0,\label{condition0}
\end{eqnarray}
which further leads to
\begin{eqnarray}
&&c_{1}=0,\nonumber\\
 &&c_{2}^{2}+2c_{3}c_{4}=0.\label{condition}
\end{eqnarray}
Eq. (\ref{condition}) is the  necessary and sufficient condition in
which an initial separable state of the system keeps in a separable
one. The nonlocal interaction between the two spins does not affect
the separability of the states in the set defined by condition
(\ref{condition}). In this case, the geometric phase of the
composite system is always equal to the sum of the geometric phases
of its subsystems.

To illustrate the above result, we consider an example. Let
$c_1=0,~c_2=-\frac{1}{\sqrt{2}}\sin\eta,~c_3=-\sin^2\frac{\eta}{2},~c_4=\cos^2\frac{\eta}{2}$,
which means that the system is initially in the stats
$\ket{\psi(0)}=\ket{00}$. At time $t$, the instantaneous state reads
\begin{eqnarray}
\ket{\psi(t)}=-\frac{1}{\sqrt{2}}\sin\eta
\ket{\psi_{2}(t)}-\frac{1}{2}
(1-\cos\eta)\ket{\psi_{3}(t)}+\frac{1}{2}(1+\cos\eta)\ket{\psi_{4}(t)},
\end{eqnarray}
where $\alpha$ and $\eta$ have been defined in Eqs. (\ref{alpha})
and (\ref{eta}), respectively. The states of the subsystems $a$ and
$b$ can be calculated by using Eq. (\ref{rhoab}), which gives
$\rho_\mu=\bordermatrix{&\cr &\rho^\mu_{11}&\rho^\mu_{12}\cr
&\rho^\mu_{21} &\rho^\mu_{22}},~~\mu=a,b, $ with the elements
\begin{eqnarray}
\rho^\mu_{11}&=&1-\rho^\mu_{22}=\frac{1}{2}\left(1+\cos^2\eta+\sin^2\eta\cos\alpha
t\right),\nonumber\\
\rho^\mu_{12}&=&{\rho^{\mu}_{21}}^*=\frac{1}{2}\left(\sin\eta\cos\eta(1-\cos\alpha
t)+i\sin\eta\sin\alpha t\right)e^{-i\omega t}.
\end{eqnarray}
By using the formulae (\ref{pure}) and (\ref{mixed}), we can
calculate the geometric phases of the system and the subsystems, and
we have
\begin{eqnarray}
\gamma_{ab}(\tau)=\arctan\frac{-2\cos\eta\sin\alpha\tau}{\sin^{2}\eta+(1+\cos^{2}\eta)\cos\alpha\tau}
+\frac{\omega\sin^{2}\eta}{\alpha}\sin\alpha\tau+\alpha\tau\cos\eta-\omega\tau\sin^{2}\eta,\label{G00}
\end{eqnarray}
and
\begin{eqnarray}
\gamma_a(\tau)=\gamma_b(\tau)=\arctan\frac{-\cos\eta\sqrt{1-\cos\alpha\tau}}
{\sqrt{1+\cos\alpha\tau}}+\frac{\omega\sin^{2}\eta}{2\alpha}\sin\alpha\tau+\frac{1}{2}\alpha\tau\cos\eta
-\frac{1}{2}\omega\tau\sin^{2}\eta.\label{g00}
\end{eqnarray}
By comparing Eq.(\ref{G00}) with Eq.(\ref{g00}) we see that the
geometric phase of the subsystem is half of the large system.

In passing, we would like to point out that all the states in the
set defined by condition (\ref{condition}) are the eigenstates of
the interaction Hamiltonian. There is no time-dependent state that
is always separable but not an eigenstate of the interaction
Hamiltonian. This is easy to be understood, since the interaction
does not change the entanglement of an eigenstate but changes that
of a non-eigenstate. Noting that the time-dependent state of the
system, initially in a separable state with $c_1\neq 0$, will be
cyclically separable with the period $t=\pi\left/4J\right.$, one may
argue whether the geometric phase holds the additivity cyclically
too in the case where $c_{2}^{2}+2c_{3}c_{4}-c_{1}^{2}=0$ but
$c_1\neq 0$. A further discussion can show that the additivity is
not valid for the geometric phase in the case. This is because that
geometric phase is equal to total phase minus dynamic phase, and
dynamic phase is not only dependent on the initial and final states
but also dependent on the states at all the evolutional time
$t\in[0,\tau]$. The additivity is invalid for dynamic phase,
although it is valid for total phase, which is only dependent on the
initial and final states. Besides, it is worth noting that the form
of condition (\ref{condition}) is based on the expression of the
basis states $\ket{\psi_k(t)}$$~(k=1,2,3,4)$ in Eq. (\ref{a3}). It
is not gauge invariant. For example, if a $\pi$ phase difference is
introduced between $\ket{\psi_3(t)}$ and $\ket{\psi_4(t)}$, the
coefficients $c_3$ and $c_4$ would acquire a relative sign and the
condition would then read $c_2^2-2c_3c_4=0$. In general, there could
be an arbitrary phase factor between $c_2^2$ and $2c_3c_4$ in Eq.
(\ref{condition}) if an alternative expression of basis states are
taken. The form of the condition depends on the choice of phase
convention between the basis states.

\section{Summarize and remarks}
An interacting bipartite system evolves into an entangled state in
general, even if it is initially in a separable state. Due to the
entanglement, the geometric phase of the system is not equal to the
sum of the geometric phases of its two subsystems. However, there
may exist a set of states in which the nonlocal interaction does not
affect the separability of the states, and the geometric phase of
the bipartite system is equal to the sum of the geometric phases of
its subsystems. By considering a well known physical model, two
interacting spin-half particles in a rotating magnetic field, we
illustrate this point. Indeed, our calculation shows that the
geometric phase of the system is not equal to the sum of the
geometric phases of the subsystems in general. They are not equal
even if the system is initially in separable states, due to the
nonlocal interaction between the subsystems. Yet, there is such a
set of states for which the nonlocal interaction does not affect the
separability of the states, and the geometric phase of the bipartite
system is always equal to the sum of the geometric phases of its
subsystems. We give a necessary and sufficient condition for an
initial separable state to remain separable.

The geometric property of the geometric phase has stimulated many
applications. It has been found that the geometric phase plays
important roles in quantum phase transition, quantum information
processing, etc. \cite{Bohm,Carollo2005,Zhu2006}. The geometric
phase shift can be fault tolerant with respect to certain types of
errors, thus several proposals using NMR, laser trapped ions, etc.
have been given to use geometric phase to construct fault-tolerant
quantum information processer \cite{Jones2000,Falci2000,Duan2001}.

The geometric phase is useful in quantum computation, but real
physical systems are usually composite and therefore the relations
among the geometric phase of the large system and those of the
subsystems are complicated. It is very difficult to control each of
the values of them. Our result shows that it is possible to make the
phases' relations simple if the initial states are properly chosen.
In this sense, our finding may be useful both in the theory itself
and in the applications of the geometric phase. The investigation on
the current bipartite system involving two spin-half particles
implies that such a kind of states
may exist in other interacting physical systems.

\section*{Acknowledgments}
This work is supported by NFS China with No.10875072 and No.
10675076. Tong acknowledges the support of the National Basic
Research Program of China (Grant No. 2009CB929400). Kwek would like
to acknowledge financial support by the National Research Foundation
\& Ministry of Education, Singapore.


\begin{thebibliography}{99}
\bibitem{tong3} If it is only required that the geometric phase of the composite
system equals the sum of those of its subsystems at some special
time $t=\tau$, there should be no difficulties in choosing initial
states. However, here we require that the geometric phase of the
composite system is aways equal to the sum of those of its
subsystems at all the time $t\in\tau$.
\end{thebibliography}
\end{document}